\documentclass[runningheads]{llncs}
\usepackage[T1]{fontenc}
\usepackage{booktabs}
\usepackage{graphicx,verbatim}
\begin{document}
\title{NAVIUS: Navigated Augmented Reality Visualization for Ureteroscopic Surgery}
\titlerunning{NAVIUS}
%
\author{Ayberk Acar\inst{1} \and
Jumanh Atoum\inst{1} \and
Peter S. Connor\inst{1} \and
Clifford Pierre\inst{2} \and
Carisa N. Lynch\inst{3} \and
Nicholas L. Kavoussi\inst{4} \and
Jie Ying Wu\inst{1}}
\authorrunning{Acar et al.}
%
\institute{Vanderbilt University, Nashville, TN, 37235, USA\\\email{ayberk.acar@vanderbilt.edu} \and
Meharry Medical College, Nashville, TN, 37208, USA
 \and
Florida International University, Miami, FL, 33199, USA \and
Vanderbilt University Medical Center, Nashville, TN, 37232, USA}



\maketitle              
\begin{abstract}
Ureteroscopy is the standard of care for diagnosing and treating kidney stones and tumors. However, current ureteroscopes have a limited field of view, requiring significant experience to adequately navigate the renal collecting system. This is evidenced by the fact that inexperienced surgeons have higher rates of missed stones. One-third of patients with residual stones require re-operation within 20 months. In order to aid surgeons to fully explore the kidney, this study presents the Navigated Augmented Reality Visualization for Ureteroscopic Surgery (NAVIUS) system. NAVIUS assists surgeons by providing 3D maps of the target anatomy, real-time scope positions, and preoperative imaging overlays. To enable real-time navigation and visualization, we integrate an electromagnetic tracker-based navigation pipeline with augmented reality visualizations. NAVIUS connects to 3D Slicer and Unity with OpenIGTLink, and uses HoloLens 2 as a holographic interface. We evaluate NAVIUS through a user study where surgeons conducted ureteroscopy on kidney phantoms with and without visual guidance. With our proposed system, we observed that surgeons explored more areas within the collecting system with NAVIUS (average 23.73\% increase), and NASA-TLX metrics were improved (up to 27.27\%). NAVIUS acts as a step towards better surgical outcomes and surgeons' experience. The codebase for the system will be available at: \url{https://github.com/vu-maple-lab/NAVIUS}.

\keywords{Endoscopy \and Ureteroscopy \and Kidney Stone Disease \and Augmented Reality \and Navigation}

\end{abstract}
\section{Introduction}
Kidney stone disease is projected to affect 25\% of the population in 30 years, decreasing a patient's quality of life and increasing the economical burden for healthcare and patients~\cite{pozdzik2024gaps}. Endoscopy of the renal collecting system (i.e., ureteroscopy), is the gold standard for minimally invasive surgical treatment of kidney stones~\cite{wason2024ureteroscopy}. However, this surgery is challenging for urologists due to the limited field of view negatively impacting navigation and kidney stone treatment ~\cite{kezer2023defining}. To overcome these challenges, recent literature focuses on solutions such as the use of 3D printed models for preoperative planning~\cite{esperto2021new}, computer vision methods for anatomical navigation~\cite{leng2024development,oliva2023orb,acar2024towards}, as well as virtual and augmented reality (VR/AR) systems~\cite{hameed2022application,detmer2017virtual}. 

In this work, we propose NAVIUS, an AR system that aids ureteroscopy by displaying a 3D virtual map created from a preoperative computed tomography (CT) scan. The system provides interactive annotations and visualizations for stones, spatial localization of the scope tip, CT scan overlays, and feedback for areas explored within the renal collecting system. To the best of our knowledge, our work is the first to demonstrate head-mounted AR visualization of real-time scope tip localization. NAVIUS can be used for preoperative planning, mapping, tracking of scope position, and keeping track of stones and fragments. 

We evaluated the benefits of NAVIUS with user studies conducted in a mock operating room setup with renal collecting system phantoms. We analyzed the effectiveness of NAVIUS through subjective surveys and objective evaluation of recorded scope trajectories. Additionally, we document our process to build realistic phantoms. We use anatomical models extracted from CT scans to ensure structural realism and colored silicone to create visual similarity. These phantoms can be used in surgical training or ex-vivo experiments, and to collect ground truth information such as tracked ureteroscopy in a controlled environment.

\textbf{Navigation Applications.} Advancements in the sensor and computer vision fields have shown the potential to enhance ureteroscopy and other minimally invasive procedures.  For example, Fu et al. developed a registration algorithm that uses structural similarities between endoscopy images and images rendered from preoperative scans to localize the endoscope~\cite{fu2023novel}. In another study, Oliva Maza et al. used ORB-SLAM 3~\cite{campos2021orb} to provide a 3D map of the organ and pose of the ureteroscope~\cite{oliva2023orb}. Similarly, Acar et al. compared structure from motion algorithms to create a 3D map, and registered this to preoperative scans to assess visited areas~\cite{acar2024towards}.

Outside of vision, Yoshida et al. focused on using the electromagnetic (EM) tracker for endoscope navigation and compared it with traditional flouroscopy based methods in several studies~\cite{yoshida2014novel,yoshida2015advantage,yoshida2019navigation}. Another study by Zhang et al. suggested the use of multiple sensors combined with a curve fitting algorithm to estimate accurate poses and bending angles~\cite{zhang2022shape}. These studies showed the feasibility and effectiveness of EM trackers in ureteroscopy; however, they were limited by surgical monitors as displays causing a decrease in 3D perception.

\textbf{Augmented Reality.}
Augmented reality technologies allowed improvements in visualization~\cite{al2020effectiveness}, planning~\cite{pose2023real}, interaction, and communication~\cite{acar2024intraoperative} for minimally invasive surgeries~\cite{qian2019review,detmer2017virtual,checcucci2024visual}. Al Janabi et al. demonstrated the feasibility of Microsoft HoloLens (Microsoft Corporation, Redmond, WA, USA) devices in an operating room (OR) simulation setting as a visualization tool, and concluded that it can improve the procedure time and performance metrics~\cite{al2020effectiveness}. Pose et al. created a real-time integration application between HoloLens 2 and 3D Slicer~\cite{fedorov20123d}, and proved the feasibility of the system in a pedicle screw placement setup. As demonstrated by these studies, the integration of augmented reality systems to OR opens up new possibilities to improve surgical outcomes. 

NAVIUS integrates navigation and AR into a single application. This represents a step towards intuitive user interfaces for navigation applications.

\begin{figure}[tbp]
\centering
\includegraphics[width=0.93\columnwidth]{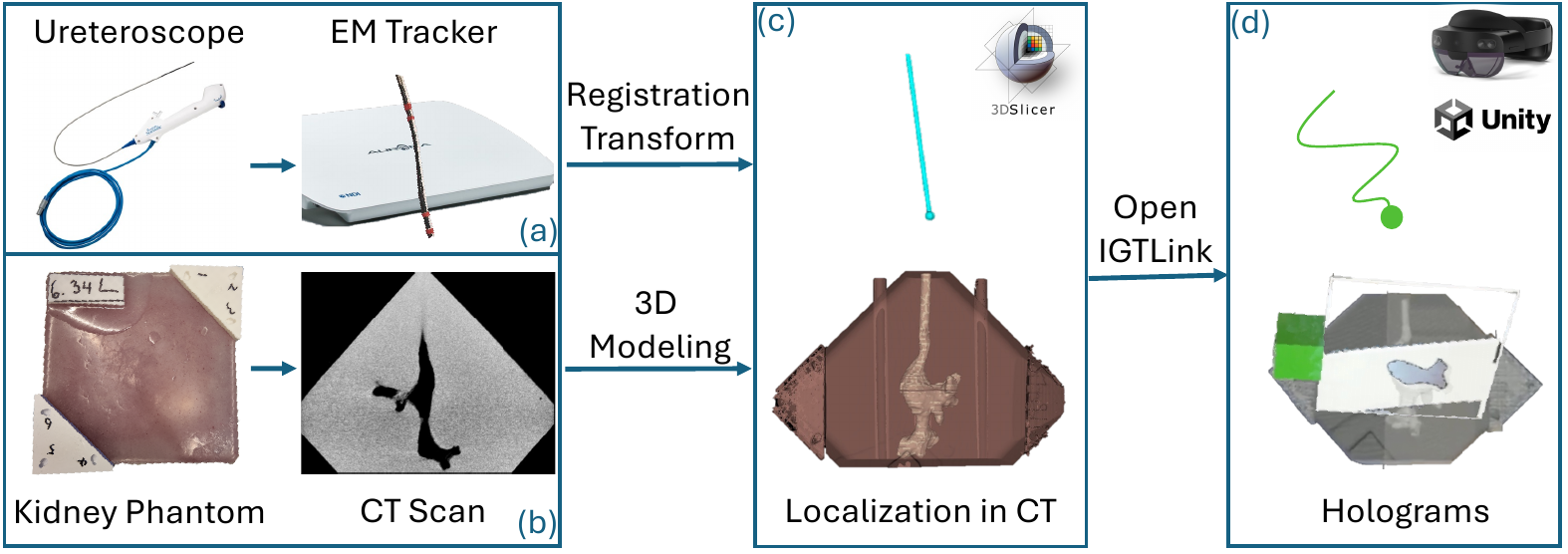}
\caption{Overall workflow of the system. We attach EM trackers to the ureteroscope and take CT scans of the phantoms. We create 3D models from CT scans and register the scope tip positions to CT coordinate plane using 3D Slicer. Scope positions and trail, CT slices, and 3D model are visualized in HoloLens 2. Letters refer to sections in the methods.}
\label{Figure-Workflow}
\end{figure}

\section{Methods}
This section explains the technical implementation and workflow of the system. NAVIUS is built upon the integration of 3D Slicer~\cite{fedorov20123d}, Unity (Unity Technologies, San Francisco, CA, USA), and HoloLens 2. An overall workflow of the system can be seen in the Fig.~\ref{Figure-Workflow}.  

\textbf{Ureteroscope Tracking.} NAVIUS tracks the position and orientation of the scope tip using an EM tracker sensor (Aurora, Northern Digital Inc. (NDI), Waterloo, ON Canada) attached rigidly to the tip, while preserving ureteroscope flexibility (Fig.~\ref{Figure-Workflow}a). 3D Slicer receives the tracking information real-time using OpenIGTLink~\cite{tokuda2009openigtlink} and PLUS toolkit~\cite{lasso2014plus}.

\textbf{Map Generation.}
We use CT scans to generate virtual 3D models of kidney. We semi-automatically segment these scans using the 3D Slicer by manually guiding modules such as ``level tracing'' and ``grow from seeds'' (Fig.~\ref{Figure-Workflow}b). In addition, stones detected in preoperative scans can be segmented using the same modules as the initial estimation for stone positions and can be added to the map. We export the segmentations as meshes in the OBJ format and import them into Unity.

\textbf{Localization in CT.}
We create a Unity scene with 3D kidney model from CT scan, a spherical game object to indicate the EM tracker position, and a user interface to control visualizations, tracking, and stone annotations. We scale the game objects, attach scripts to enable holographic interactions, and adjust the hierarchy between them to make hologram movements consistent between objects. The AR application starts by establishing a connection between Unity and the 3D Slicer on a remote computer. 3D Slicer sends the tracker position and 2D CT slice information as in~\cite{pose2023real} to Unity (Fig.~\ref{Figure-Workflow}c).

\begin{figure}[tbp]
\centering
\includegraphics[width=0.8\columnwidth]{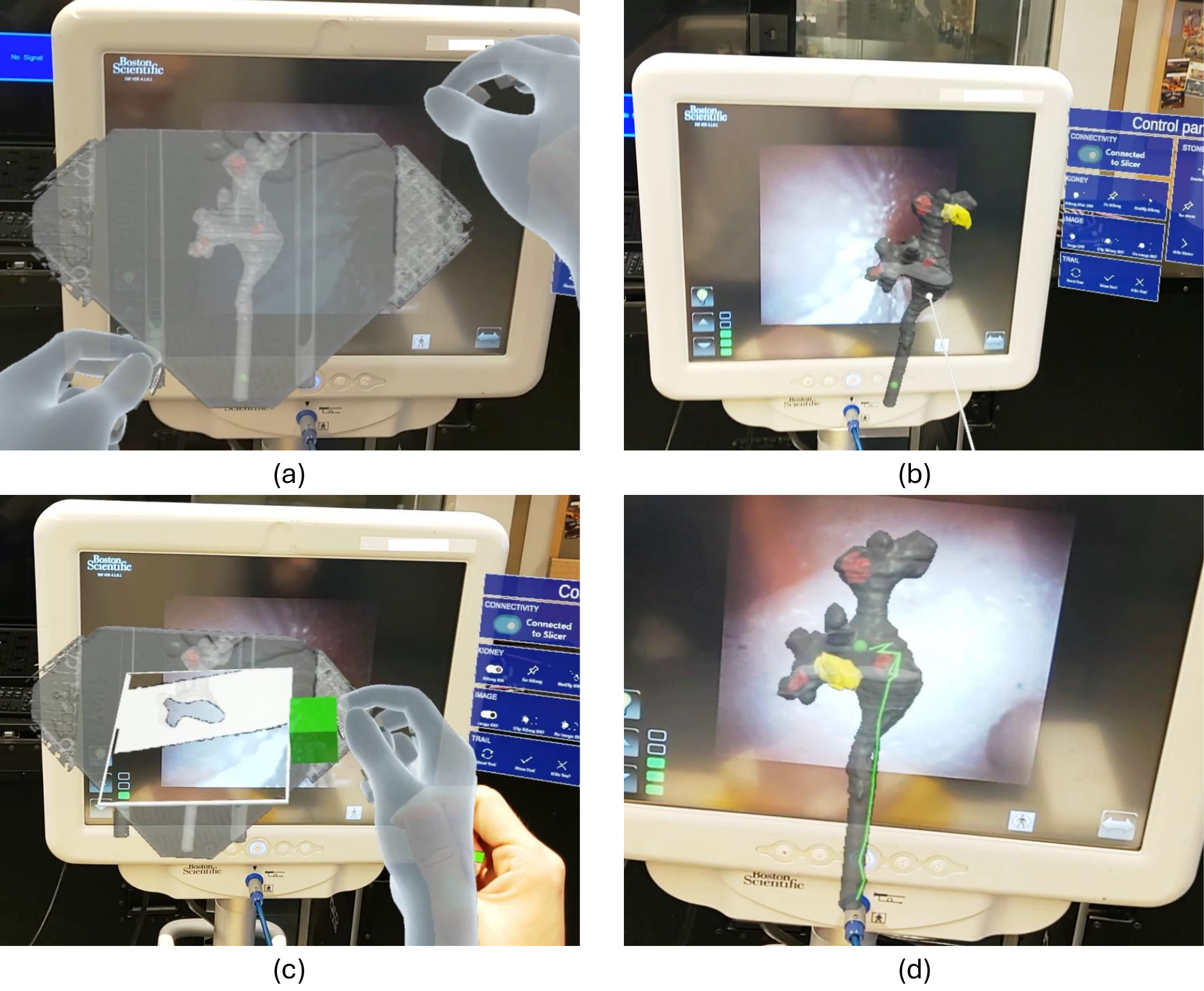}
\caption{Application features. (a) Complete anatomy visualizations that can be placed anywhere. (b) Alternatively, segmented collecting system can be used as visualization. Stones or other pathologies detected can be annotated with holographic objects of different colors. (c) Overlay of preoperative scans can be seen on holograms. (d) Scope tip position and tracked trajectory are visualized with a green marker and trail.}
\label{ApplicationFeatures}
\end{figure}

\textbf{Holograms.}
In Unity, the 3D Slicer information and Mixed Reality Toolkit features are integrated to create holograms of CT layovers and tracker positions. Those holograms are displayed on the HoloLens through holographic remoting (Fig.~\ref{Figure-Workflow}d). Using the HoloLens 2, users can place the holograms anywhere without disrupting their workflow and can see the targets from different viewpoints. A toggle between the complete anatomy visible in the CT field of view and the segmented collecting system allows surgeons to focus on the target area while retaining their spatial awareness (Fig.~\ref{ApplicationFeatures}a-b). Whenever a stone is detected, they can annotate the area with a stone hologram and choose a different color for each annotation (Fig.~\ref{ApplicationFeatures}b). Furthermore, they can visualize the 2D slices of CT scans as an overlay or as a 2D image in user interface by freely moving the image plane over the 3D model (Fig.~\ref{ApplicationFeatures}c)~\cite{pose2023real}. Finally, the application displays the scope position within the renal collecting system in real-time and creates a trail of the movement. This allows surgeons to localize the scope within the collecting system with ease, and keep track of the explored areas (Fig.~\ref{ApplicationFeatures}d). 

\section{Experiments and Results}
To evaluate the effectiveness of the proposed application, we conducted a user study with surgeons using kidney phantom models. We designed an experiment to simulate real operating workflow, where six surgeons explore the renal collecting system with and without the NAVIUS system. 

\begin{figure}[tbp]
\centering
\includegraphics[width=0.93\columnwidth]{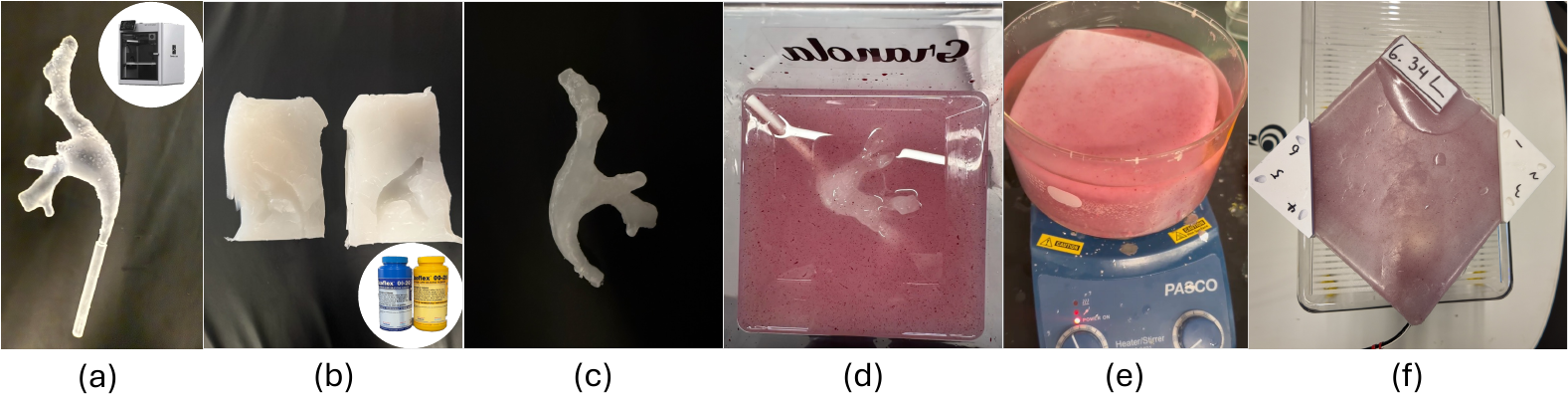}
\caption{(a) 3D printed collecting system. (b) Negative mold created with Ecoflex. (c) Wax model using mold. (d) Outer mold filling. (e) Boiling to remove wax model and create calyces. (f) Final phantom.}
\label{PhantomProduction}
\end{figure}

\subsection{Experimental Setup}
\textbf{Phantom Production.} 
To acquire anatomically accurate phantoms for experimental setup, we follow a similar methodology to Adams et al.~\cite{adams2017soft}, while eliminating the need of wax-compatible 3D printing. A detailed video explanation of the process can be found in the video: \url{https://youtu.be/Ei3HvCMJSOw}. 

We use renal collecting system meshes segmented from patient CT scans and print with a Bambu Lab X1C 3D printer (Bambulab USA Inc, Austin, TX, USA) (Fig.~\ref{PhantomProduction}a). Using a cylindrical container, we mix 500g each of Ecoflex 00-20 silicone rubber (Smooth-On, Inc., Macungie, PA, USA) parts A and B, degas, and submerge the model, ensuring central positioning. A weighted plate keeps it submerged during solidification at room temperature for four hours.

After curing, we remove the Ecoflex mold and cut it into equal halves. This Ecoflex mold serves as the outer cast for the wax inner mold (Fig.~\ref{PhantomProduction}b). We melt paraffin wax at 70$^{\circ}$C and pour it into the ecoflex cast via the ureter inlet. We carefully extract the wax mold after it solidifies at room temperature within an hour (Fig.~\ref{PhantomProduction}c).

To create the external mold, we place the wax inner mold in a square container, mix 250g each of Ecoflex parts A and B with one tablespoon of red food coloring, degas for 1 minute, and pour over the mold while ensuring central positioning. As the wax mold partially floats, we allow initial curing for two hours before adding another Ecoflex layer (Fig.~\ref{PhantomProduction}d). After four hours, we remove the extender piece and submerge the phantom in 70$^{\circ}$C isopropyl alcohol or ethanol to dissolve the wax (Fig.~\ref{PhantomProduction}e).  Finally, we rinse, dry, and perform an endoscopic evaluation to confirm complete dissolution of wax. For our validation study, we use two phantoms that are evaluated as having similar exploration difficulty by an attending surgeon.

\begin{figure}[tbp]
\centering
\includegraphics[width=0.95\columnwidth]{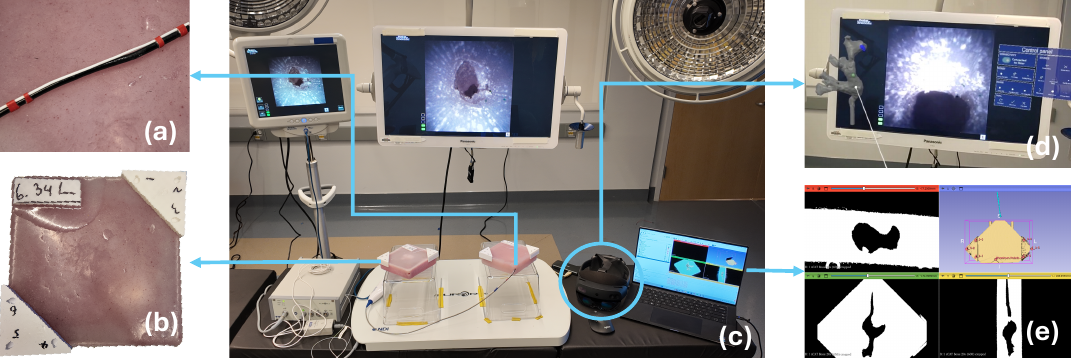}
\caption{Experimental setup (a) EM tracker attachment to scope. (b) Phantoms with registration fiducials. (c) Mock OR experimental setup. (d) HoloLens 2 field of view. (e) 3D Slicer view with registered scope positions and preoperative scans.}
\label{Figure-ExpSetup}
\end{figure}

\textbf{Ureteroscope Tracking.}
During the experiments, we attached an EM tracker to a Boston Scientific (Boston Scientific Corporation, Marlborough, MA, USA) single-use flexible ureteroscope using circular heat shrinks (Fig.~\ref{Figure-ExpSetup}a) This allows a rigid attachment to the tip without restraining the scope movement or significantly increasing the diameter. We fixed the phantoms rigidly on a tabletop field generator using double sided tapes. For the calibration, we 3D printed pieces with three scope tip slots and attached two of those rigidly to the phantoms (Fig.~\ref{Figure-ExpSetup}b). At the beginning of the experiments, we collected six data points by moving the scope to 3D printed reference slots, and using the ``Fiducial Registration Wizard'' within SlicerIGT module~\cite{ungi2016open}, registered the tracker and CT scan coordinate planes.

\textbf{Experiments.}
Six surgeons with different levels of experience with HoloLens 2 and ureteroscopy explored the renal collecting systems. Each surgeon conducted two trials in two different conditions in different phantoms: with application (WA) and no application (NA). All surgeons reviewed the preoperative CT scans of each phantom before starting each procedure. In the WA condition, participants used the AR system with HoloLens 2, with a 3D model of the phantom and real-time localization. In the NA condition, they followed the standard workflow of preoperative CT scan review (i.e., prior to the procedure) and thereby relied on their spatial memory during the operation. We randomized the order of phantoms and the availability of the AR application to minimize the learning effects at each trial. Participants completed the NASA Task Load Index (NASA-TLX)~\cite{hart1988development} to assess subjective workload and the System Usability Scale (SUS)~\cite{brooke1996sus} to evaluate the AR application. In addition, we recorded the real-time tracking data and quantitatively evaluated the explored area and distance traversed with the scope. All experiments were approved by institutional review board.

\subsection{Results}
\textbf{NASA-TLX Results.}
To measure the cognitive load and surgeons' experience with the application, participants were asked to complete NASA TLX survey after each trial, rating 6 different categories from 1 to 20 (Table~\ref{NASA_TLX_table}). 

\begin{table}[tbp]
\centering
\caption{NASA-TLX Metric results. Values given are average assessments of 6 users. Lower is better. Best values are indicated in bold.}
\begin{tabular}{c|c|c|c}
\toprule
                         & \textbf{No Application} & \textbf{With Application} & \textbf{\% Change} \\
\hline
\textbf{Mental Demand}   & 13.33                   & \textbf{11.00}            & -17.50             \\
\hline
\textbf{Physical Demand} & \textbf{9.83}           & 10.17                     & 3.39               \\
\hline
\textbf{Temporal Demand} & \textbf{7.00}           & 7.50                      & 7.14               \\
\hline
\textbf{Performance}     & 9.17                    & \textbf{6.67}             & -27.27             \\
\hline
\textbf{Effort}          & 13.00                   & \textbf{10.17}            & -21.79             \\
\hline
\textbf{Frustration}     & 10.67                   & \textbf{7.83}             & -26.56         \\
\hline
\end{tabular}
\label{NASA_TLX_table}
\end{table}

\textbf{SUS Results.} Participants are asked to fill a questionnaire with 10 questions by ranking the statements from 1 (strongly disagree) to 5 (strongly agree). Using the SUS calculation method~\cite{brooke1996sus}, we calculated a final score for each user. We observed the same average SUS score of 67.08 on average of 6 users, for both the baseline NA scenario and the WA scenario. In our results, only two questions had >1 difference in average responses for the two scenarios while the other metrics show no significant differences. Question 4 on the need for support (negative effect on the score) was rated 1.83 for NA and 3.83 for WA. Question 7 on ease of learning (positive effect) was rated 2.67 for NA and 4.17 for WA.

\begin{table}[bp]
\centering
\caption{Quantitative results acquired by endoscopy trajectory. Values given are averages of 4 users with complete trajectory recordings during the exploration.}
\begin{tabular}{c|c|c|c}
\toprule
                        & \textbf{No Application} & \textbf{With Application}  & \textbf{\% Change}\\
\hline
\textbf{Convex Hull Volume $(\uparrow)$}      &10603.42 $mm^3$          &13119.28 $mm^3$ &23.73\\
\hline
\textbf{Total Distance Traveled} &757.25 $mm$ &747.50 $mm$       & 1.30\\
\hline
\end{tabular}\label{QuantitativeResults}
\end{table}

\textbf{Quantitative Results.} 
To evaluate the performance quantitatively, we analyzed the EM tracker data recorded with Perk Tutor Transform Recorder module for 3D Slicer~\cite{ungi2012perk}. We applied an outlier removal to the trajectories to discard unrelated points created during the retraction. We used a Mahalanobis distance~\cite{mahalanobis2018generalized} threshold determined by trial and error based on trajectory visualizations. With a threshold of 3.0, 3.57\% of points were considered outliers. For volume coverage analysis, we fit a convex hull over the trajectories (Table~\ref{QuantitativeResults}). 

\section{Discussion and Conclusion}
\begin{figure}[tbp]
\centering
\includegraphics[width=0.55\columnwidth]{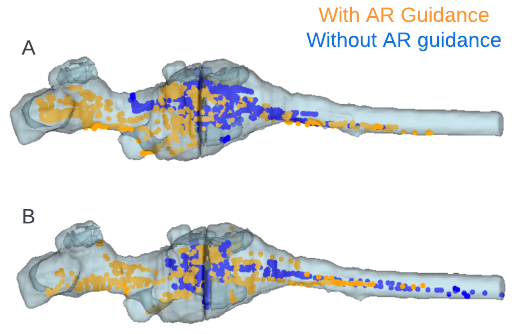}
\caption{Trajectories resulting from AR guidance (in orange) show in-depth exploration of the kidney phantom as opposed to without AR guidance (in blue). (A) and (B) shows trajectories created by four users exploring the kidney phantoms; 2 with AR guidance and 2 without.}
\label{Trajectories}
\end{figure}

NAVIUS results in improvements in the four NASA-TLX metrics and in the total volume explored. Using the AR application led to the greatest improvement in performance (27.27\%) and reduced frustration (26.56\%), as shown by the NASA-TLX (Table~\ref{NASA_TLX_table}). Additionally, when the users are asked if they have any additional comments, without any prompting, 3 out of 6 surgeons stated NAVIUS would boost their confidence in exploring the kidney. Quantitative results support this, as surgeons covered 23.73\% more volume on average without significantly increasing scope travel distance (Table~\ref{QuantitativeResults}, Figure~\ref{Trajectories}). We report results from four uses since the trajectory collection on two users was incomplete. This increase in volume coverage indicates that with the help of NAVIUS, surgeons can do a more thorough exploration and avoid missed areas or stones.

We observe increases in physical and temporal demand metrics of NASA-TLX survey but these changes are less significant compared to improved metrics. These limitations may be caused by the surgeons' unfamiliarity with the system and the added stimulation from the application. User responses to SUS question 4 about the need for technical support reinforce this hypothesis. These limitations can be improved as surgeons become more familiar with HoloLens 2 and NAVIUS. In question 7 of SUS, participants claimed this application is easier to learn compared to regular ureteroscopy flow. 

NAVIUS scored 67.08 ($\pm$ 7.65) in SUS evaluation, which is around the general average and accepted benchmark of 68 ($\pm$ 12.5)~\cite{hyzy2022system}. The same average SUS scores for the procedure with and without the proposed AR application show that the application does not introduce significant additional challenges to usability.

Integration of the system to the OR can be affected by factors such as decreased EM tracker performance due to interference by other devices and the need to attach the sensor to the scope. However, usability can be improved with additional calibration procedures~\cite{lugez2015electromagnetic}, and our future work includes the replacement of EM trackers with visual navigation methods. 

To conclude, we introduce NAVIUS, an AR application to aid surgeons during kidney stone surgeries. In the phantom studies, we observe improvements in both NASA-TLX metrics and the volume covered within the collecting system. This system can be extended to other surgical areas, and can be integrated with robotic systems using interfaces such as SlicerROS2~\cite{connolly2024slicerros2}. Future work can focus on introducing additional functionalities such as collaborative virtual environments and integration of endoscopic segmentations. Overall, this application can improve the operational outcomes for the patient and surgeons' experience in the OR. NAVIUS demonstrates the feasibility of navigation and AR integration to transform ureteroscopy workflow.

\bibliographystyle{splncs04}
\bibliography{bibliography}
\end{document}